\documentclass[lettersize,journal]{IEEEtran}
\usepackage{amsmath,amsfonts}
\usepackage{algorithmic}
\usepackage{array}
\usepackage[caption=false,font=normalsize,labelfont=sf,textfont=sf]{subfig}
\usepackage{textcomp}
\usepackage{stfloats}
\usepackage{url}
\usepackage{verbatim}
\usepackage{graphicx}
\def\BibTeX{{\rm B\kern-.05em{\sc i\kern-.025em b}\kern-.08em
    T\kern-.1667em\lower.7ex\hbox{E}\kern-.125emX}}
\usepackage{balance}

\usepackage[ruled]{algorithm2e}
\usepackage{tcolorbox}
\usepackage{multirow}
\usepackage{booktabs} 
\newcommand{\answer}[2] {
    \vspace{-1ex}
	\begin{tcolorbox}[boxrule=1pt,left=1pt,right=1pt,top=2pt,bottom=2pt]
		\textbf{Answer to RQ#1:} #2
	\end{tcolorbox}
	\vspace{-1ex}
}
\usepackage{tikz}
\usepackage[multiple]{footmisc}
\usepackage{makecell}

\usepackage{pifont}
\tikzset{
    cross/.pic = {
        \draw[rotate = 45] (-#1,0) -- (#1,0);
        \draw[rotate = 45] (0,-#1) -- (0,#1);
    }
}

\newcommand\mynuma[1]{\ifcase#1 \or \ding{172}\or \ding{173}\or
	\ding{174}\or \ding{175}\or \ding{176}\or \ding{177}%
	\or \ding{178}\or \ding{179}\or \ding{180}\or \ding{181}\else *\fi\relax}

\newcommand\mynumb[1]{\ifcase#1 \or \ding{182}\or \ding{183}\or
	\ding{184}\or \ding{185}\or \ding{186}\or \ding{187}%
	\or \ding{188}\or \ding{189}\or \ding{190}\or \ding{191}\else *\fi\relax}

\newcommand\mynumr[1]{\ifcase#1 \or \romannumeral1\or \romannumeral2\or
	\romannumeral3\or \romannumeral4\or \romannumeral5\or \romannumeral6%
	\or \romannumeral7\or \romannumeral8\or \romannumeral9\or \romannumeral10\else *\fi\relax}

\begin{document}

\title{Demystifying the DAO Governance Process}

\author{Junjie Ma, Muhui Jiang,  Jinan Jiang, Xiapu Luo, Yufeng Hu, Yajin Zhou, Qi Wang, Fengwei Zhang}

\maketitle
\begin{abstract}
Decentralized Autonomous Organization (DAO) becomes a popular governance solution for decentralized applications (dApps) to achieve decentralized governance. In the DAO, no single entity can arbitrarily control the dApps without approval from the majority of members. However, despite its advantages, DAO has also been targeted by several attacks, leading to the loss of millions of dollars. 
In this paper, we first provided an overview of the DAO governance process within the blockchain. Next, we identified the issues within three components of governance process: Governance Contract, Documentation, and Proposal. Each of these components is vulnerable to issues that could potentially result in substantial financial losses. Then we developed automated methods to detected above issues. To investigate the issues within the existing DAO ecosystem, we constructed a state-of-the-art dataset that includes 16,427 DAOs, 183 documentation, and 122,307 proposals across 9 different blockchains. Our analysis reveals that a majority of DAO developers and members have not given sufficient attention to these issues, especially in the area of proposal. The result shows that over 60\% of the examined proposals fail to provide a consistent description and code for their members, highlighting a significant gap in ensuring transparency within the DAO governance process. For a better DAO governance ecosystem, DAO developers and members can utilize the methods to identify and address issues within governance process.

\end{abstract}


\section{Introduction}

\IEEEPARstart{D}{ecentralized} Autonomous Organization (DAO) is a governance method constructed based on blockchain smart contracts~\cite{ETH_WP}. 
The DAO ensures that all privileged actions must gain the majority of member consensus, thus effectively preventing arbitrary actions from a certain member.

Recently, a significant number of decentralized applications (dApps) have adopted DAO as their governance method. For example, Uniswap~\cite{uniswap}, one of the most valuable Decentralized Exchange (DEX), with a daily trading volume exceeding 500 million dollars~\cite{CoinMarketCap}, employs DAO for its asset management. Additionally, DAO platforms such as XDAO~\cite{xdao_Doc}, Aragon~\cite{aragon}, and DAOhaus~\cite{daohaus}, which help developers to deploy DAO in minutes, have attracted the interest of thousands of organizations~\cite{Platform_based_DAO}. In particular, XDAO~\cite{xdao_Doc}  has facilitated the setup of over 33,000 DAOs across various blockchains. According to the analysis~\cite{deepDAO}, the total treasury governed by DAOs exceeds 18.8 billion dollars, with over 2.5 million users participating in DAO governance. This trend indicates that DAO has become a popular governance method adopted by blockchain developers.

However, this rapid increase in DAOs has come with challenges. Many DAO developers, as well as members, fail to pay sufficient attention to the issues in the DAO governance process. This oversight has led to an increase in attacks targeting DAOs~\cite{yam_attack, curve_attack, audius_attack, Fortress_protocol_attack, create2_attack, True_Seigniorage_Dollar_Attack, Pride_punks_attack, buildfinance_attck, yuan_attack, beanstalk_exploit}. For instance, the governance process can be attacked due to the absence of clear explanations describing the code to be executed in the proposal. An example is the Beanstalk attack, leading to a loss of 182 million dollars~\cite{beanstalk_exploit}. The attacker deceived members into trusting the code in the proposal was benign. In reality, the code intended to transfer all assets owned by Beanstalk to the attacker. Moreover, DAOs can also be manipulated by the developers. Normally, the dApps' token contract should be controlled by the governance process, ensuring that without the majority's permission, no one can transfer the locked tokens in the contract. However, in some DAOs, the contracts are controlled by specific developers. This allows the developers to arbitrarily control the contract without obtaining permission from the governance process. An example of this is the VPANDA DAO Rug Pull~\cite{VPANDA_DAO_rug_pull}, where the developer illegally transferred over 1 million locked tokens from the token contract to swap for over 265 thousand dollars, resulting in a 99\% drop in the token value.

Previous studies within the field of DAOs have predominantly focused on analyzing DAO activities and issues related to voting in the governance process~\cite{dao_gas_price,DAO_overview,vulnerableDAO,rw_voting, empirical_on-chain,empirical_practice,DAO-voting-platform, empirical_voting, novel_on_chain}, such as centralized voting power. To the best of our knowledge, there is no work focusing on the issues of the entire DAO governance process. Our work fills this gap by conducting a comprehensive study and answering the following 3 research questions. Each research question is related to a distinct aspect within the governance process identified in the section~\ref{process}. 

\noindent\textbullet \textbf{RQ1}: Do existing DAOs achieve fairness decentralized governance?

\noindent\textbullet \textbf{RQ2}: Do existing DAOs offer sufficient governance process documentation for their members?

\noindent\textbullet \textbf{RQ3}: Do existing proposals ensure consistency between descriptions and code?

For RQ1, we verify that the DAO achieves decentralized governance, ensuring developers can not compromise the fairness of the governance process. We first adopt static analysis of the governance contract to ascertain whether it has the required governance functions. Then, we extract the controller addresses of privileged functions to determine whether the governance contract is self-governed or controlled by developers. Lastly, we trace the creation process of the governance contract to ensure that developers can not arbitrarily modify the contract's code logic.
For RQ2, we investigate whether the DAO offers adequate guidance to its members for participating in the governance process, thereby motivating member engagement in governance process. We employ Large Language Model (\textbf{LLM}) with Chain of Thought (\textbf{CoT})~\cite{gpt-chain} to evaluate if the DAO documentation complies with the six requirements outlined in the DAO Model Law~\cite{model_law}.
In RQ3, we assess whether the proposals submitted by members exhibit consistent and immutable code behavior, aligning with their descriptions.  Thus, attackers cannot disguise malicious proposals as legitimate ones to misappropriate the DAO's funds. Initially, we trace the proposal code to verify its immutability after submission. Subsequently, we employ a combination of Natural Language Processing (\textbf{NLP}) and \textbf{LLM} to ensure that all actions prescribed by the code are accurately reflected in the proposal descriptions.
The issues pertaining to DAOs, as discussed in our study, have not been previously explored in other research. Furthermore, our investigation encompasses an extensive collection of over 16,000 DAOs from 5 distinct platforms and  9 different blockchains.

Our results show that not all DAO governance processes are precisely implemented. We identify one DAO in which the governance contract can be destructed and redeployed at the same address by the developer.
In terms of providing documentation to assist members in participating in the governance process, over 98\% of DAOs fail to provide such documentation. Given that such documentation is intended to equip members with essential information for DAO governance, its absence can deter member participation in governance or lead to controversial voting results.
For these proposals in the governance process, we find that only 34\% of proposals (8,584) contain descriptions about the code actions in the proposal. Furthermore, of these proposals, less than 30\% (2,536 proposals) provide a complete explanation of the code, such as what function will be invoked and how many tokens will be transferred. To assess the effectiveness of our approach in detecting real-world governance attack cases, we evaluate our approach against the latest 13 governance attack incidents. Our approach successfully detects all of these attacks.
We hope our paper can guide developers in deploying and maintaining their DAOs in a more comprehensive and secure manner and also enlighten members about potential risks within the DAO governance.

Our contributions can be summarized as follows.



\noindent\textbullet \textbf{Public Dataset:} We collected 16,427 different DAO implementations, 183 documentation, and 122,307 proposals across 9 popular blockchains. Our dataset included famous DAOs such as Uniswap~\cite{uniswap} and Compound~\cite{compound_doc}, as well as DAOs from platforms like Aragon~\cite{aragon}. The collected data will be released for further research.

\noindent\textbullet \textbf{Comprehensive Study:} 
We conducted a comprehensive study on a large amount of DAO implementations, gaining an in-depth understanding of the issues addressed in the 3 proposed research questions concerning the DAO governance process. Our findings revealed that over 99\% of DAOs failed to provide documentation. Besides, over 90\% of the existing proposals failed to elucidate the code actions.




\noindent\textbullet \textbf{Insightful Findings:} We found the current DAO implementations have many security issues, which deserve our attention. 1) More than 600 DAOs contain privileged functions controlled by unidentified entities, potentially serving as backdoors that could compromise the governance process. 2) The governance contract code logic can be arbitrarily modified without changing its address, allowing developers to manipulate governance results.

\section{Background}
\subsection{Decentralized Autonomous Organization}
Decentralized Autonomous Organization (DAO) in blockchain is first introduced by Ethereum white paper~\cite{ETH_WP}. DAO utilizes smart contracts to enable collective control of the organization by all its members. Using smart contracts in the DAO allows for establishing organization rules and managing the treasury through immutable code. Currently, there are two types of DAO governance~\cite{two_governance_type}\cite{two_governance_type-2}: on-chain governance and off-chain governance. On-chain governance requires all the governance processes to be conducted on the blockchain by smart contracts, including proposing proposals, voting, and executing. On the contrary, in off-chain governance, the decision-making process (e.g., proposing proposals or voting) is performed outside the blockchain. The execution process is carried out manually by the DAO developer, granting it complete control over the DAO contracts. We exclude off-chain governance from our scope as off-chain governance contravenes the requirements in the DAO definition~\cite{ETH_WP} and the DAO Model Law~\cite{model_law}, which mandate governance to be executed on smart contracts.

\subsection{DAO Platform} The DAO platform is designed to provide DAO developers with the tools to easily create their own DAOs. Developing a DAO requires advanced programming and blockchain knowledge. Current DAO platforms such as XDAO~\cite{xdao_Doc}, Aragon~\cite{aragon}, DAOhaus~\cite{daohaus}, and DAOstack~\cite{daohaus} offer comprehensive assistance in DAO creation. Their assistance spans a wide range, from on-chain smart contracts deployment to the creation of a dedicated voting website. In these cases, developers can create their own DAO in minutes. 

\subsection{DAO Model Law}\label{back}
The DAO Model Law~\cite{model_law} is a type of Model Law~\cite{the_model_law}, seeks to bridge the divide between DAOs and traditional regulatory frameworks yet to adapt to new company structures fostered by the blockchain. DAO Model Law stipulates rules applicable to both the on-chain smart contracts and off-chain documentation. Once these rules are met, DAOs and their members can achieve legal certainty. The DAO Model Law is the only document that provides rules for smart contracts and documentation.

\section{ DAO Governance Process}\label{process}
\begin{figure}

\begin{center}
\includegraphics[width=1 \columnwidth]{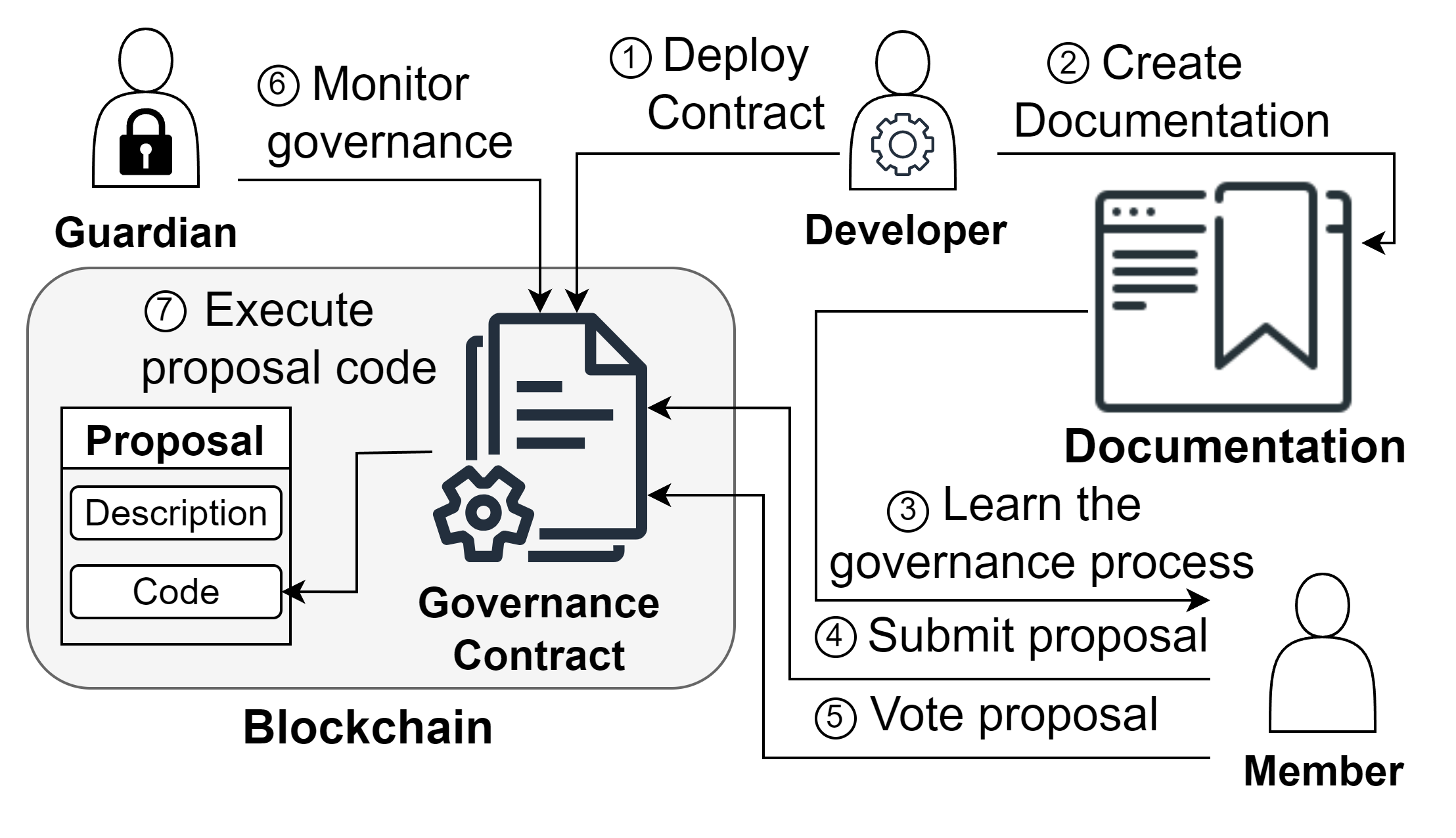}
\end{center}

\caption{ The DAO Governance Process.}\label{fig:DAO_summary}

\end{figure}

We provide a comprehensive overview of the DAO governance process, as shown in Figure~\ref{fig:DAO_summary}.

\noindent\textbf{Participants.} The participants in the DAO governance process fall into one of three roles: \emph{developer}, \emph{member}, and \emph{guardian}. The first role, \emph{developer}, is involved in the development of the DAO's smart contracts and interface. He is responsible for \ding{172} deploying the governance contracts to the blockchain network, as well as \ding{173} creating the documentation for the DAO. The second role, \emph{member}, is a blockchain user who learns the governance process by \ding{174} reading the documentation. He can participate in DAO governance by \ding{175} submitting or \ding{176} voting for a proposal via the DAO governance contract~\cite{compound_doc}. The last role is the \emph{guardian}, a specific blockchain user tasked with \ding{177} monitoring the DAO governance process within the governance contract. If the \emph{guardian} detects malicious proposals targeting DAO governance, he has the authority within the governance contract to cancel such proposals.


\noindent\textbf{Governance contract.} The governance contract controls the governance process, storing all the proposals and votes from members. It provides functions that allow members to submit new proposals, vote on these proposals, and execute the code within the proposals. The governance contract should be configured as the only way to change the parameters of DAO contracts.


\noindent\textbf{Proposal.} Proposal refers to a formal submission to governance contract that is made by a member to suggest changes to the DAO (i.e.,  funding request, contract parameters configuration). Typically, as shown in Figure~\ref{fig:CA}, the proposal encompasses two elements: \emph{description} and \emph{code}.
The \emph{description}, penned in natural language, outlines the intent of the proposal. It provides members with information regarding the proposal code as well as the reason behind it.
The \emph{code} contains the code that will be executed by governance contract if the proposal gets passed. It refers to the technical implementation of the proposal.

\noindent\textbf{Governance process.} Managing and implementing changes within a DAO relies on the governance process. This is achieved by submitting proposals to the governance contract and conducting votes on these proposals. If a proposal passes the voting process, the code within it is executed by the governance contract to implement the changes towards the DAO. This ensures all the changes are approved by the majority of the DAO members. The governance process begins at a \emph{member} \ding{175} submitting a proposal to the governance contract. Then, a \emph{member} can \ding{176} cast vote for the newly submitted proposal. A proposal is passed when it has received sufficient voting power in support from members. If the \emph{guardian} does not identify the proposal as a malicious one \ding{177}, the code within the proposal will be executed by the governance contract \ding{178}.

\noindent\textbf{Documentation.} Besides the governance contracts, the DAO also needs to provide the \emph{documentation}. 
Considering the complexity of the governance process, the \emph{documentation} should provide complete guidance on governance process. This encompasses delivering detailed information on how to become a DAO member, providing step-by-step guides to participate in the governance process, and outlining the existence of \emph{guardian}.
\section{Approach}

\subsection{Research Questions}\label{RQ}

We examine the issues within each component of the governance process - \textbf{Governance Contract}, \textbf{Documentation}, and \textbf{Proposal} -with the following research questions.

\begin{figure}
\begin{center}
\includegraphics[width=1 \columnwidth]{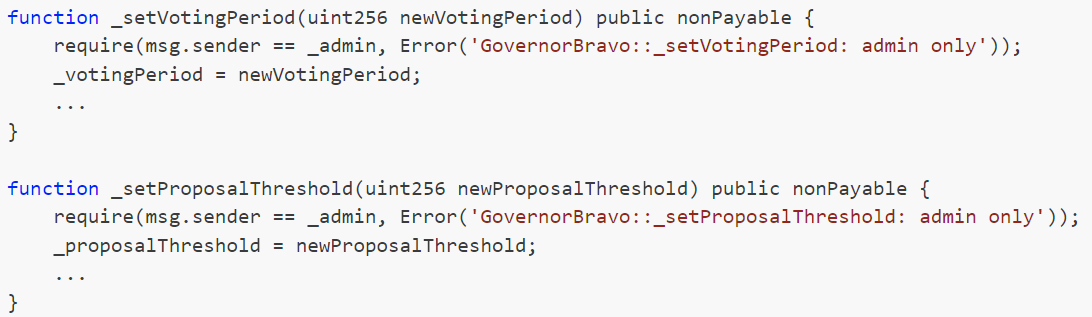}
\end{center}
\caption{The decompiled governance contract from \textbf{mini dao} shows that the developer controls privileged functions (\emph{setVotingPeriod} and \emph{setProposalThreshold}), enabling him to control proposal voting duration and required voting power.}\label{fig:RQ1_example}
\end{figure}

The \textbf{Governance Contract} controls the entire governance process. Therefore, according to the definition~\cite{ETH_WP}, it is essential for achieving fairness in decentralized governance, preventing developers from arbitrarily manipulating the results. 
However, a malicious developer could embed privileged functions within the governance contract that are controlled by themselves. This manipulation allows them to undermine the governance process, compromising the fairness of proposal outcomes. For instance, functions such as \emph{setVotingPeriod} and \emph{setProposalThreshold} within the governance contract are designed to set the proposal voting duration and the required voting power for a proposal to pass. Normally, these functions can only be called by the governance contract itself, ensuring that only passed proposals can modify these parameters. However, in the case of the governance contract 0x41E6......7a42 from the DAO mini dao shown in Figure~\ref{fig:RQ1_example}, these functions are controlled by an \emph{admin}, which is an Externally Owned Account (\textbf{EOA}) specified by the developer, rather than the governance contract itself. Thus, the developer can pass any proposal by adjusting the voting delay to allow only themselves to vote, or cancel any proposal by setting the proposal threshold to a high value.

Thus, we propose RQ1 to examine whether the existing Governance Contracts belonging to DAOs achieve comprehensive in decentralized governance.

\textbf{RQ1:} Do existing DAOs achieve fairness decentralized governance?

\begin{figure}
\begin{center}
\includegraphics[width=1 \columnwidth]{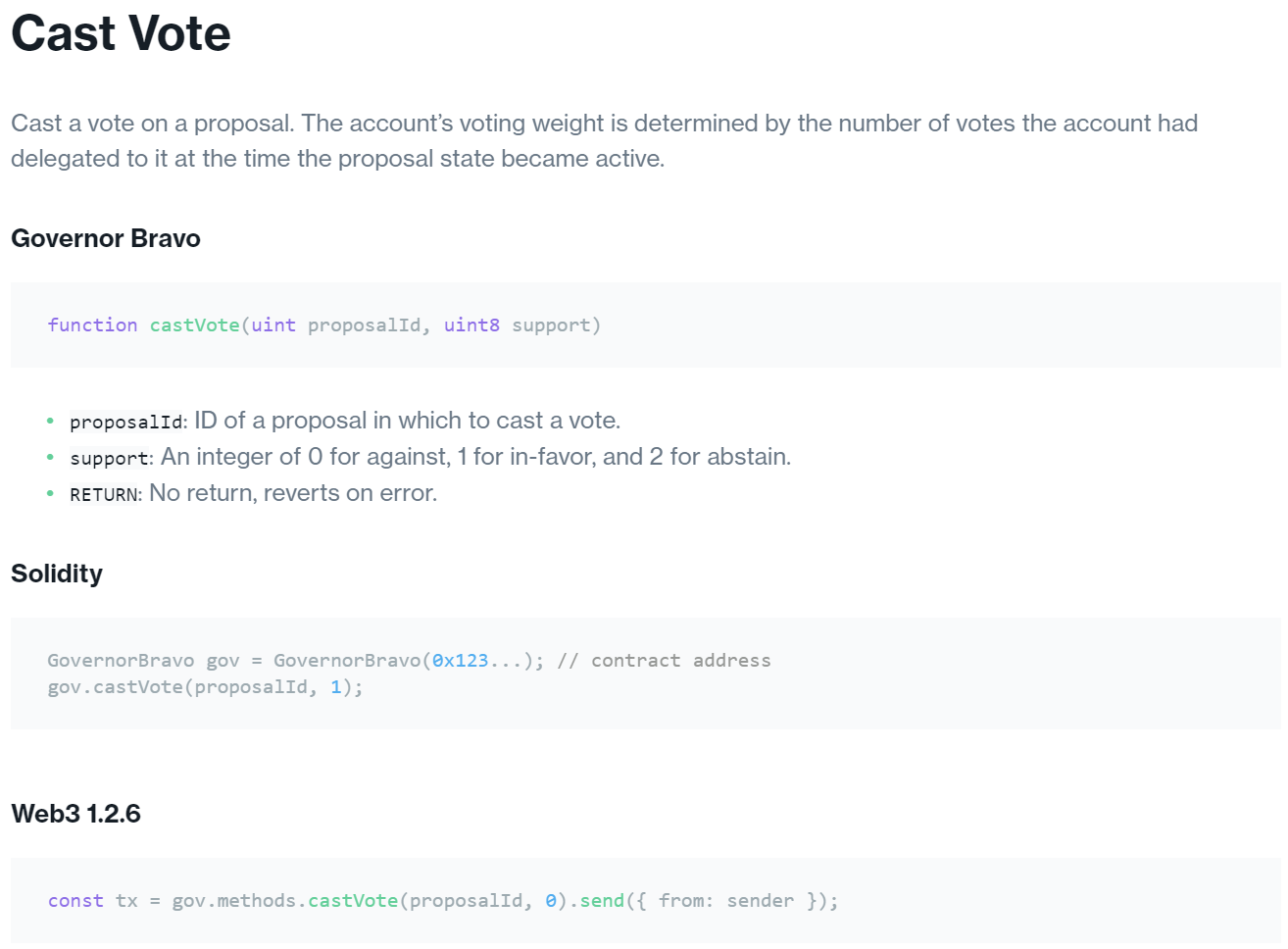}
\end{center}
\caption{The Compound governance documentation provides DAO members with guidance on how to vote for proposals.}\label{fig:Compound_doc}
\end{figure}

As for the \textbf{Documentation}, each DAO should provide detailed documentation for their member instructs members on participating in the governance process, emphasizing the disclosure of its critical aspects.
For example, as illustrated in Figure~\ref{fig:Compound_doc}, the DAO Compound offers comprehensive documentation for its members, detailing how to engage in the governance process.
The lack of governance documentation can hinder members engagement in governance, as members need to read the governance source code to learn how to participate in the governance process. This scenario could result in a situation where governance result is determined and controlled by only a handful of members. For example, in the Synthetify DAO governance attack on 17 October 2023~\cite{Synthetify}, an attacker submitted a malicious proposal aiming to seize control of the DAO's assets. Due to the lack of governance documentation, none of the DAO members actively participated in the governance process, and as a result, no one vetoed this malicious proposal during the 7-day voting period. This oversight led to a loss of 230 thousand dollars.

We assess in RQ2 whether DAOs provide sufficient documentation on governance processes for their members.

\textbf{RQ2:} Do existing DAOs offer sufficient governance process documentation for their members?

\begin{figure}
\begin{center}
\includegraphics[width=1 \columnwidth]{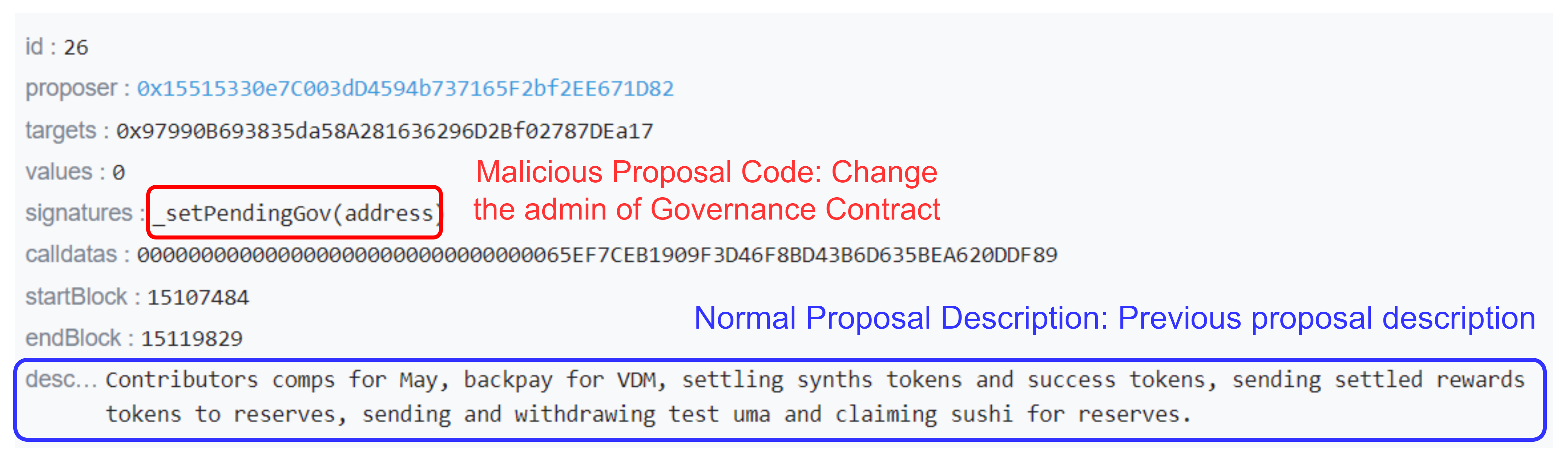}
\end{center}
\caption{The malicious proposal in the YAM governance attack deceived members with a description from a previous proposal, claiming it would return rewards to the DAO. However, the actual intent of the code was to take control of the governance contract.}\label{fig:YAM}
\end{figure}

As for the \textbf{Proposal}, attackers can submit malicious proposals to gain control of the DAO or misappropriate the DAO's assets by embedding malicious code within the proposal. They may deceive members into believing the code is legitimate by providing a misleading description. For example, during the governance attack on the DAO YAM on July 9, 2022~\cite{yam_attack}, the attacker proposed a proposal shown in Figure~\ref{fig:YAM}. The attacker misled members with a description from a previous proposal, claiming it aimed to return rewards to the DAO. However, the code actually transferred ownership of the governance contract from the DAO to the attacker, resulting in a loss of 3.1 million dollars once it succeeded.

In RQ3, we examine the consistency between the proposal descriptions and the code to prevent malicious members from submitting proposals that deceive other members by disguising malicious intentions as legitimate actions.

\textbf{RQ3:} Do existing proposals ensure consistency between descriptions and code?

\subsection{Data Collection}
In this section, we collect data related to DAOs to address the research questions mentioned above. The result of our dataset is listed in Table~\ref{tab:DAO_dataset}. Specifically, we collect DAO information, websites, documentation, and proposals. However, the collection of these data presents several challenges.
First, there is a lack of comprehensive dataset that encompasses all information related to DAOs.
Second, current DAO information does not include DAO documentation and only provides a few DAO websites.
Third, not all DAO platforms offer APIs to fetch proposals. 
To solve these challenges, we outline the method for each type of data as follows.

\noindent\textbf{DAO information.} We locate the DAO information mainly from previous studies~\cite{dao-analyzer, DAO-voting-platform} and available DAO datasets. Specifically, we gather DAO information from 1 DAO dataset, DeepDAO~\cite{deepDAO}, and 5 DAO platforms (i.e., XDAO~\cite{xdao_Doc}, Aragon~\cite{aragon}, DAOhaus~\cite{daohaus}, DAOstack~\cite{daostack}, and Tally~\cite{tally}). Note that our data focuses on EVM-compatible chains whose Total Value Locked (TVL) is larger than 50 million dollars.  We collect DAO information from these sources until Feb 1, 2024.

To filter out unused DAOs from these DAO sources, which might cause bias in the overall trustworthy results, we exclude DAOs that have never proposed any proposals.

We acquire the DAO name (or id), governance contract address, and website (if available). Since data collected from DeepDAO includes DAOs also present in the other platforms, we remove DAOs from DeepDAO that are duplicated in other data sources. Note that two DAOs with identical contract addresses are treated as the same one.  Finally, we collect 3 distinct DAOs from DeepDAO, which are not included in the other 5 platforms and are marked as independent by DeepDAO. We label these 3 DAOs as self-developed. To ensure our dataset is complete enough, we verify its contents with the top 20 DAOs listed on CoinMarketCap\cite{CoinMarketCap}. The result shows that all the top 20 DAOs, including Uniswap\cite{uniswap} and Compound\cite{compound_doc} from Tally, as well as Curve\cite{curve} and MakerDAO\cite{makerDAO} from DeepDAO, are included.

\begin{table}[]
\renewcommand{\arraystretch}{0.85}
\caption{ Types of DAOs, along with their corresponding quantity, website, documentation, and proposal in the database.}\label{tab:DAO_dataset}
\resizebox{0.99\columnwidth}{!}{
\begin{tabular}{@{}lrr||rr||r@{}}
\toprule
 DAO Type & DAO (total) &  DAO (used) & Website & Documentation & Proposal \\ \midrule
XDAO & 33,685&  14,415 & 336& 38 & 97,156 \\
Aragon & 2,384& 1,182 & 111& 44 & 16,149 \\
Tally & 615 & 544 & 144 &92 & 3,953 \\
DAOhaus & 243 & 221  & 7&1 & 1,712 \\
DAOstack & 219 & 62 & 21&5 & 2,190 \\
Self-developed & 3 & 3 &3& 3 & 1,047 \\ \midrule
Total & 37,149 & 16,427 & 622 & 183 & 122,207 \\ \bottomrule
\end{tabular}
}
\end{table}

\noindent\textbf{Documentation.} Collecting the corresponding documentation is not easy as the aforementioned dataset usually do not provide such resources directly.
To address this, our initial step involves searching for the DAO website within the dataset or its associated DAO platform. Should this approach prove unsuccessful, we try to query its public name tag~\cite{name_tag}  from blockchain scanner to find whether the governance contract has linked to the DAO website. Then, we utilize Selenium~\cite{Selenium} to crawl through the DAO website to gather the documentation. Specifically, we focus on links containing the keywords $whitepaper$ or $doc$. If such specific links are not found, we archive the entire website.

\noindent\textbf{Proposal.} To retrieve the proposals, we utilize the interfaces provided by platforms such as Aragon, DAOhaus, and DAOstack to download all the proposals corresponding to each DAO. If these platforms do not provide the proposal information for the DAO, we retrieve the event logs from the DAO's governance contract address and extract the proposal information inside the logs.



\section{Do existing DAOs achieve fairness decentralized governance?(RQ1)}\label{RQ1}

In this section, we examine whether existing governance contracts have implemented fairness in decentralized governance. Specifically, we assess three aspects of the governance contract: soundness, independence, and immutability. 
For soundness, we verify whether the governance contract has the capability to achieve the intended governance processes. The absence of such capability constitutes a violation of the requirements of a DAO. 
Secondly, we examine independence by ensuring that the governance process is controlled by the governance contract. This ensures that developers are prevented from compromising governance outcomes by invoking privileged functions. 
Lastly, regarding immutability, we investigate whether the governance contract's code logic can be altered by developers, which could lead to manipulation of governance and misappropriation of assets from members.

\subsection{Soundness of Governance Contract} 
As stipulated by the DAO definition~\cite{ETH_WP, model_law}, the governance of a DAO must be decentralized. 
This necessitates that the DAO achieves decentralized governance through its governance contract. Therefore, if the governance contract lacks the capability to facilitate decentralized governance, it constitutes a violation of the fundamental principles of a DAO.

\noindent \textbf{Approach.} 
To evaluate whether a DAO has soundly implemented decentralized governance within its Governance Contract, we employ different methods for different types of DAOs.

For DAOs from platforms XDAO, Aragon, DAOhaus, and DAOstack, it is mandated that they utilize the template governance contracts provided by the respective platforms~\cite{DAO_overview}. 
Initially, we conduct a manual analysis to verify whether the contracts provided by these platforms soundly implement decentralized governance. Subsequently, we ascertain whether a DAO belonging to the aforementioned types adopts the provided governance contract. 
Therefore, if the governance contracts provided by the platform facilitate decentralized governance and the corresponding DAOs adopt the same governance contract as provided by the platform, we determine that such a DAO achieves the decentralized governance.
To verify whether a DAO's governance contract matches the template governance contracts, we first trace the creator of the DAO governance contract by obtaining the governance contract creation transaction from the corresponding blockchain. If the creator address of the DAO governance contract is identified as the deployer address listed in the platform's deployment guidance, we confirm that the governance contract is the same as the template provided by the platform. For those governance contracts whose creators differ from the platform deployer, we download the bytecode of the DAO governance contract and the corresponding template contract from the respective blockchain. We then compare them to ascertain whether they are identical.

For DAOs from Tally, the developers are allowed to add new functions based on the template contract provided by OpenZeppelin~\cite{proposal_code_description} or Compound~\cite{compound_doc}. We can not directly compare the bytecode of these contracts to ascertain if it is the same as the template contract. We check whether the governance contract includes the three governance functions from the template contract (i.e., \emph{Propose}, \emph{Vote}, and \emph{Execute}) as required by the DAO Model Law~\cite{model_law} as well as the template contract from OpenZeppelin and Compound. 
(1)\emph{Propose.} A member can submit a proposal by invoking this function. 
(2)\emph{Vote.}  For a proposal recorded in the contract, members have the ability to cast their votes using this function. 
(3)\emph{Execute.} The function can execute the code of the proposal. 
If the DAO's governance contract includes all three aforementioned functions, we deduce that it adheres to the template contract. To determine whether the target governance contract possesses the required functions, we compute the similarity between the functions in the governance contract and the required functions.

For the rest of DAOs, if the governance contract is open-source and the documentation supports decentralized governance while also providing its address, we infer that such a DAO has achieved decentralized governance. Otherwise, we check whether the governance contract is similar to the contract provided by the platform or contains similar functions from the template contract.

To compute the similarity between contracts, we follow the methodology proposed by previous study~\cite{ngram}. Specifically, we extract the bytecode of the contracts and eliminate the parameters of opcode $PUSH$. Therefore, we compute hypervectors of n-grams (n=5) of opcodes for each contract. The similarity score is calculated by the Jaccard similarity of their respective hypervectors. If the similarity score exceeds the threshold, we determine that the two contracts are similar. We adhere to prior research by setting the threshold value at 0.8~\cite{similarity}.

To determine the similarity between functions, we utilize EVM CFG BUILDER~\cite{evm-cfg-builder} to extract each function's bytecode from the contract. We adopt the same method used for computing the similarity scores between contracts to compute the similarity between functions. In order to mitigate the effect of different Solidity versions causing discrepancies in the bytecode pattern, we compile the contracts with each major Solidity version. If the target function matches any version of the template contract function, we determine the functions are similar.

\noindent \textbf{Result.} The results, as shown in Table~\ref{tab:DAO_governace_type}, reveal that all the used DAOs achieve decentralized governance.

To verify the effectiveness of our approach, we extend it to those DAOs that do not have proposals. Among 20,722 unused DAOs, we detect 303 DAOs that do not achieve decentralized governance, including 193 from Aragon and 109 from DAOstack. We further conduct a manual analysis on these 303 DAOs to identify any false positives and determine their causes. Within the 194 DAOs associated with the Aragon platform, we find that they do not provide the governance contract address. This issue is attributed to the DAO developers' decision to omit decentralized governance support during the creation process via Aragon. Similarly, for the 109 DAOs on the DAOstack platform, we discover that they do not support decentralized governance due to their original design. There do not exist any false positives.
These observations underscore that DAO platforms provide developers with the discretion to either incorporate or exclude decentralized governance during DAO creation. To adhere the principles of decentralization, DAO platforms might consider making decentralized governance support a mandatory feature for developers.

\begin{table}[!t]
\centering
\renewcommand{\arraystretch}{0.85}
\caption{Numbers of DAOs that achieve decentralized governance ($DG$), along with those where privileged functions in the governance contract are controlled by the governance contract or other entities.}\label{tab:DAO_governace_type}
\renewcommand{\arraystretch}{0.9}
\begin{tabular}{@{}lrr||rr@{}}
\toprule
DAO Type        &  With $DG$& Without $DG$  & Governance& Other\\ \midrule
XDAO            & 14,415    & 0             & 14,296    & 119\\
Aragon          & 1,182     & 0           & 775     & 407\\
Tally           & 544       & 0             & 515       & 29\\
DAOhaus         & 221       & 0             & 221       & 0\\
DAOstack        & 62       & 0          & 62       & 0\\
Self-developed  & 3         & 0             & 3         & 0\\ \midrule
Total           & 16,427   & 0           & 15,872    & 555 \\
\bottomrule
\end{tabular}%
\end{table}



\subsection{Independence of Governance Contract}
\begin{figure}
\begin{center}
\includegraphics[width=1.0 \columnwidth]{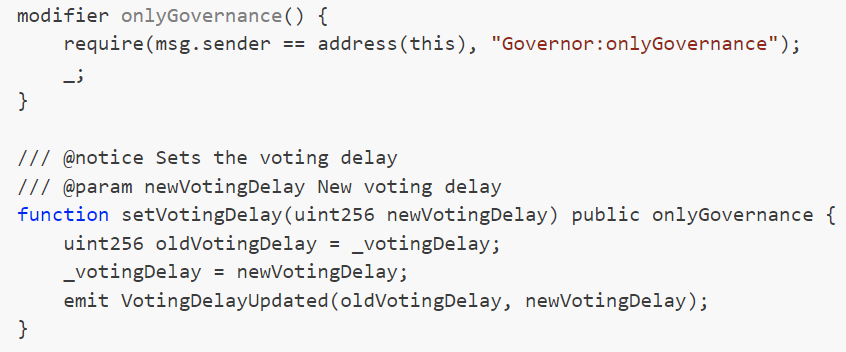}
\end{center}
\caption{A simplified privilege function restriction requires the function caller to be the governance contract.} \label{fig:func_caller}
\end{figure}

All privileged functions within the governance contract should be controlled by the governance contract itself to ensure the governance process is independent and protected from any potential violations by developers.
The privileged function is defined as a function that can be executed only by a privileged address~\cite{caller, caller2}. As shown in Figure~\ref{fig:func_caller}, the $setVotingDelay$ function, designed to adjust the voting time within the governance contract, has a modifier named $onlyGovernance$. This modifier specifies that only when the function caller's address $msg.sender$ matches the governance contract itself, the function can be invoked, indicating that this function is under the governing of the governance contract. However, if certain DAO functions are not governed by the governance contract, this could lead to security vulnerabilities, as illustrated in Figure~\ref{fig:RQ1_example}. An attacker could manipulate the governance process by controlling the outcome of any proposal. This could be achieved by arbitrarily adjusting the voting period and the required voting power needed to pass the proposal.



\noindent \textbf{Approach.} For DAOs from platforms XDAO and Aragon, these DAOs adopt the same contract for their governance logic and access control. Meanwhile, these platforms also provide an official  API~\cite{xdao_Doc}\cite{aragon} to query the governor of the privileged functions. Thus, we can determine whether the governance contract governs all privileged functions by querying with the provided APIs.

For the others, inspired by previous studies~\cite{caller}\cite{caller2}, we apply static analysis of the governance contract bytecode to identify privileged functions and extract the privileged address from these functions. In particular, to identify privileged functions, we investigate whether a function requires the caller's address, obtained from the opcode $CALLER$, to compare with a specific address from contract storage by the opcode $EQ$. The result is used to determine the $jump$ target. We then extract this address and compare it with the governance contract address to ascertain if they align.

\noindent \textbf{Result.} As demonstrated in Table~\ref{tab:DAO_governace_type}, the majority of DAOs, particularly those on platforms like DAOhaus and DAOstack, strictly adhere the requirement that all functions within the governance contract should be governed by the governance contract itself.

However, 555 governance contracts retain certain privileged functions that are not governed by themselves. As indicated by the results in Section~\ref{RQ2}, most DAOs fail to explain the existence of guardians. Thus, it is hard for members to classify whether these functions are potentially backdoors or designated for guardians to protect the governance process.

\subsection{Immutability of Governance Contract}\label{lab:RQ1_3}


After the Constantinople update~\cite{Constantinople}, the EVM introduced a new opcode, $CREATE2$, which allows a smart contract to be deployed at a predetermined address~\cite{create2}. This can be exploited as an attack vector since it allows contract developers to modify the contract code after deployment while maintaining its address unchanged~\cite{create2_attack,create2_attack_source}. Compared to the normal contract creation opcode $CREATE$, where the contract address is computed as $keccak256(address+nonce)$, $address$ represents the contract creator's address, and $nonce$ denotes the total transaction number of the contract creator. In $CREATE2$, the contract address is calculated as $keccak256(0xff+address+salt+keccak256(CODE))$. Here, $address$ refers to the contract creator's address, $salt$ is a 32-byte string, and $CODE$ corresponds to the init code of the new contract. This introduces a potential risk, as it permits developers to arbitrarily modify the contract code while maintaining the same contract address by the $CREATE2$. However, the predetermined contract address from $CREATE2$ can not already have a contract code. This limitation can be circumvented using the opcode $SELFDESTRUCT$, which can destroy a contract along with its code.

Thus, the attack can be launched with the following steps. An attacker $X$ first utilizes $CREATE2$ to deploy contract $A$ at the address calculated as $ADDR(A)==keccak256(0xff+ADDR(X)+0x0+keccak256(CODE(A)))$. Thus, $X$ uses contract $A$ to deploy the governance contract $G$ using the opcode $CREATE$ at the address as $ADDR(G)==keccak256(ADDR(A)||0)$. To change the code of the governance contract $G$, the attacker first destroys contracts $G$ and $A$ using the $SELFDESTRUCT$ opcode. The attacker then deploys a new contract $A'$ by $CREATE2$, using the same salt and contract code as $A$. Given that $A'$ and $A$ share identical salt and contract code, the address of $A'$ will be the same as that of $A$: $ADDR(A')==keccak256(0xff+ADDR(X)+0x0+keccak256(CODE(A)))==ADDR(A)$. In the final stage, the attacker deploys the new governance contract $G'$ using $A'$ and the $CREATE$ opcode. As $A'$ is a newly deployed contract, its nonce remains 0, leading the new governance contract $G'$ to have the same contract address as $G$: $ADDR(G')==keccak256(ADDR(A)||0)==ADDR(G)$. Thus, despite being created by $CREATE$, contract $G'$ has the same address as $G$, but with a different code. Therefore, it is essential to ensure the governance contract is not at risk of arbitrary modifications to its contract code via $CREATE2$.

\noindent \textbf{Approach.} We first define the Contract Creation Chain (CCC) of a governance contract as follows: Given a governance contract address $G$, we trace its contract deployment transaction. If it is deployed by a contract $C_0$, we add it to the CCC. We then trace the creator of $C_0$, designated as $C_1$, and continue this process until we find a contract that is created by an EOA address $E$. The $CCC(G)= <G, C_0, C_1,......,E>$ shows the governance contract $G$ is created from a chain of contracts that extend from $C_0$ to $E$.

The procedure to determine whether a governance contract is at risk from $CREATE2$ 
is 
shown below.
For a given governance contract address $G$, we first construct its CCC. To do this, we use the corresponding blockchain scanner to query the respective contract deployment transaction. After building the CCC, For each contract $C_i$ in $CCC(G)$, we first investigate whether the contract can destruct itself via the opcode $SELFDESTRUCT$ to erase its own code. However, a potential attacker could conceal the opcode within a different contract and indirectly execute it using the $DELEGATECALL$ to destroy the original contract. Hence, if a contract contains the opcode $SELFDESTRUCT$ or $DELEGATECALL$, we conclude that it can destruct itself. Subsequently, in order to check whether contract $C_i$ is created by $CREATE2$, we trace the opcodes used during the contract deployment transaction. If the $CREATE2$ opcode is used to create $C_i$, we deem that $C_i$ is created by $CREATE2$. We take advantage of the Tenderly API~\cite{Tenderly} to access the sequence of executed opcodes from the deployment transaction of contract $C_i$. 
Finally, if we determine that contract $C_i$ is created by $CREATE2$ and all preceding contracts within the chain can destruct themselves, our algorithm returns true; otherwise, returns false.

\noindent \textbf{Result.} We identified one DAO from Tally, associated with the governance contract $0xe28...d5$. This contract is created by $CREATE2$ and utilizes the $DELEGATECALL$ opcode to interface with external contracts. Notably, such contracts can be destroyed by developers and subsequently redeployed at an identical address.

\answer{1}{Among the 16,427 DAOs analyzed,  665 could be manipulated by developers. Besides,  DAOs listed by DAO platforms can not be considered trustworthy, as these platforms do not mandate the actual achievement of decentralized governance and lack the mechanisms to verify the consistency of a DAO's governance contract with its intended template contract. Furthermore, there exists one DAO whose governance contract code logic can be manipulated by the developer using the $CREATE2$ opcode.}

\section{Do existing DAOs offer sufficient governance process documentation for their members?(RQ2)}\label{RQ2}


The documentation is expected to provide a complete overview of the DAO, elaborating on the governance process interactions for its members. Considering that the DAO Model Law\cite{the_model_law}, as referenced in Section~\ref{back}, prescribes specific rules for documentation, we have consolidated the documentation requirements from the DAO Model Law into the following six rules.
1) \emph{Member Participation.} 
The documentation should provide guidelines on how blockchain users can become DAO members and participate in governance, as well as the participation rights in the governance process.
2) \emph{Member Exit.} Apart from participating in DAO, the documentation should also describe the steps a member needs to follow to exit the DAO, whether in a voluntary or involuntary way.
3) \emph{Voting Power.} The documentation should clearly explain how voting power is calculated and distributed among members, as voting power determines the weight of a member's vote. Failing to explain voting power could discourage member participation in voting or, conversely, enable a member to accumulate excessive voting power, potentially allowing him to arbitrarily control the result of voting.
4) \emph{Minority Protection.} The documentation should explicitly state if it includes any provisions for protecting the minority rights of its members. This is crucial because minority members may need to raise disputes against specific decisions, particularly in situations where a single member controls the majority of voting power.
5) \emph{Governance Process Guide.} A detailed guide to the governance process is necessary for members. For instance, the step-by-step instructions for submitting proposals and casting votes.
6) \emph{Appointment of Guardian.} The appointment of a guardian is crucial to alleviating security concerns among members. Given the significant privileges the guardian holds, such as controlling the privilege functions in the governance contract, their role should be disclosed in the documentation. 


Considering most of the members are not able to reliably and accurately extract information from the on-chain DAO contract code, it's vital that the DAO presents this information in the transparent, publicly accessible document. The absence of such transparency may erode members' trust, thereby discouraging their active participation in DAO governance.


\noindent \textbf{Approach.} We employ ChatGPT~\cite{chatgpt} as a question-answering system to determine whether the above 6 rules are present in the DAO documentation. Based on recent studies~\cite{gpt_quesiton_reason0}\cite{gpt_quesiton_reason2}\cite{gptvsbert}\cite{gpt_quesiton_reason1}, ChatGPT surpasses existing Large Language Model  (LLM) models in terms of performance in such question-answering task. Furthermore, it exhibits superior robustness in question comprehension when compared with state-of-art question-answering systems.

\begin{figure}
\begin{center}
\includegraphics[width=0.9 \columnwidth]{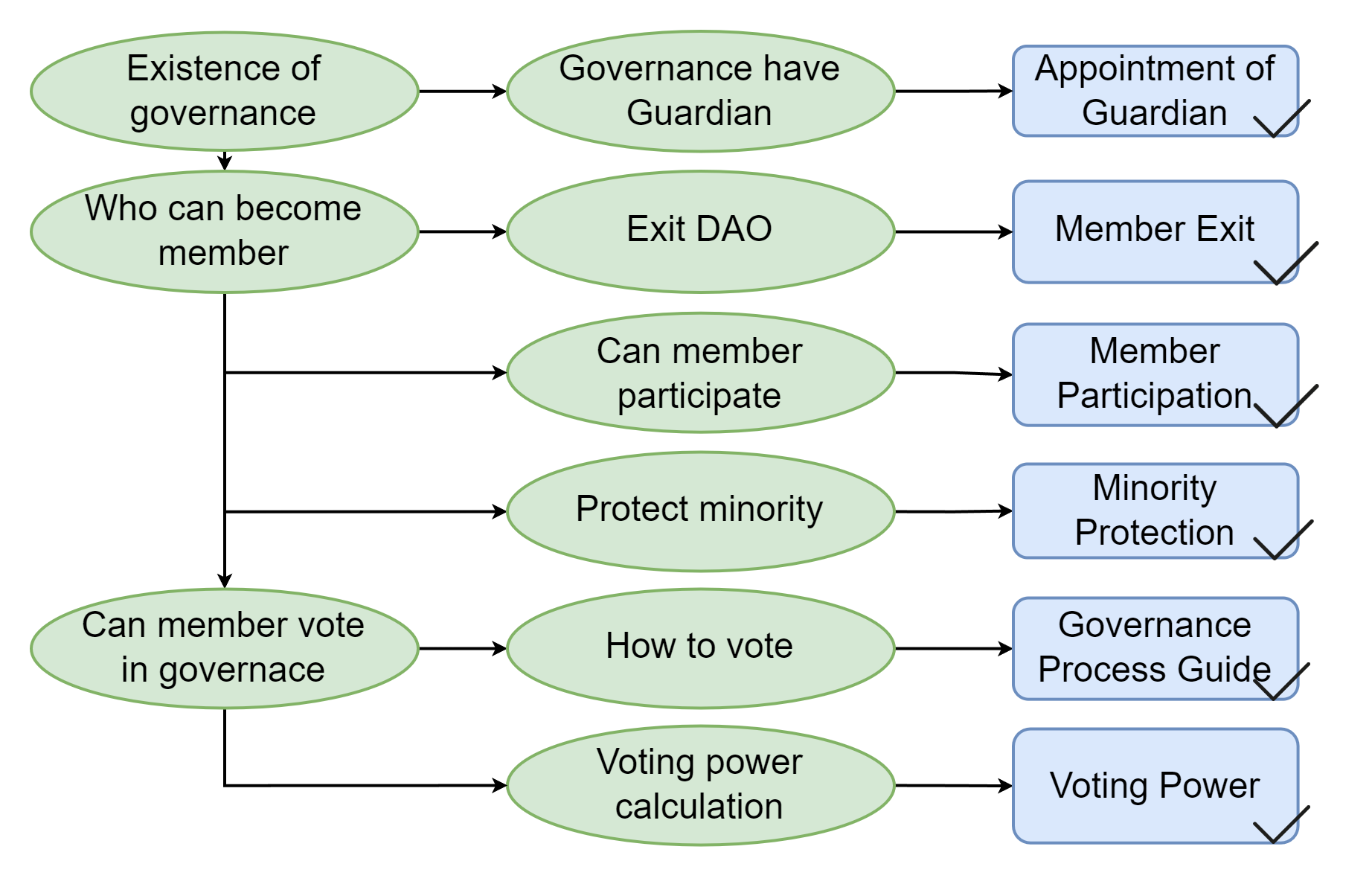}
\end{center}
\caption{The abbreviated question chain to query whether the 6 rules are mentioned in the documentation. Each arrow represents a \emph{Yes} response from ChatGPT.}\label{fig:QA_chain}
\end{figure}

\begin{figure}
\begin{center}
\includegraphics[width=1 \columnwidth]{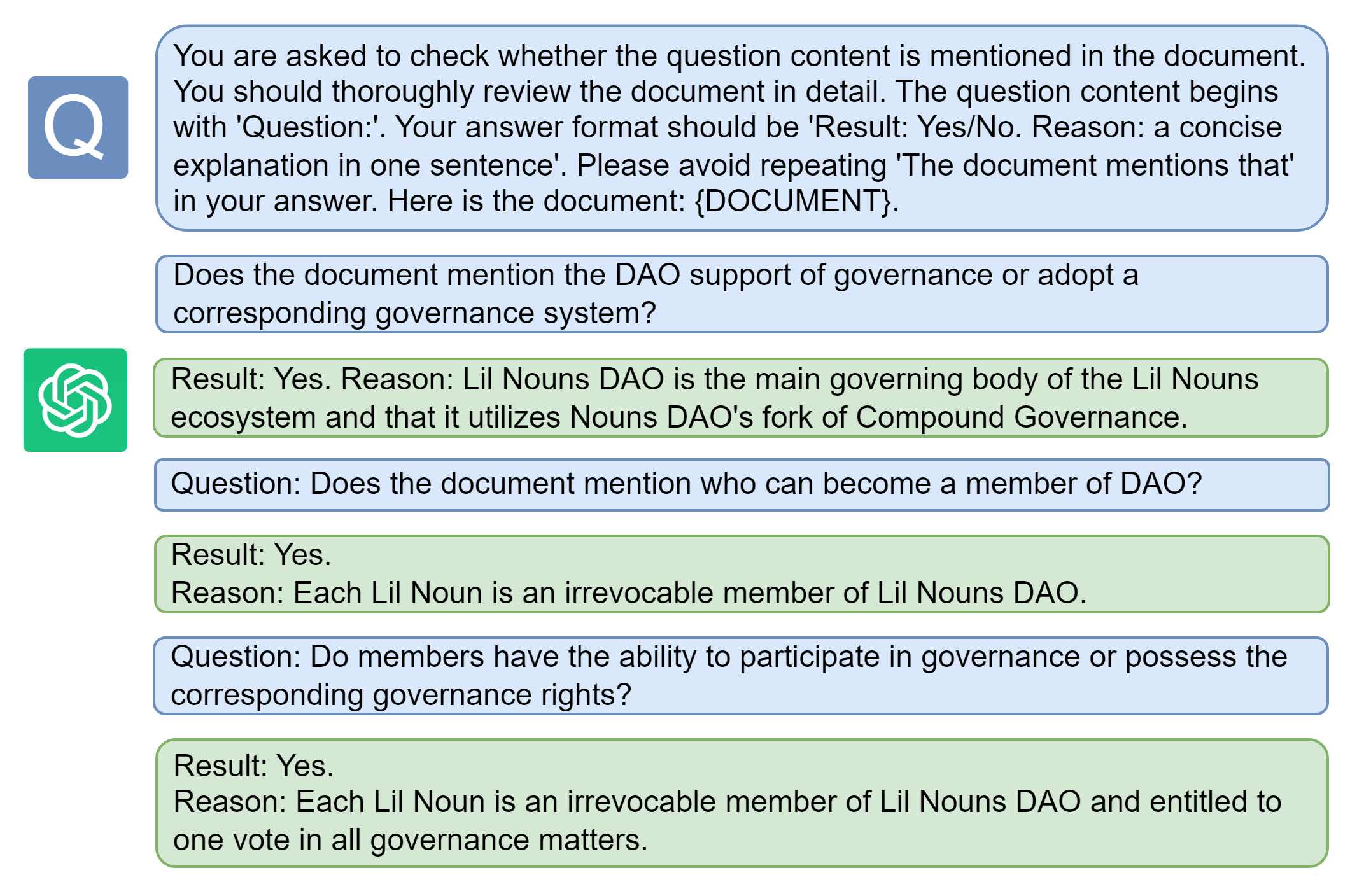}
\end{center}
\caption{An example of querying ChatGPT with a chain of questions to check the Rule1 \emph{Member Participation} is mentioned in the documentation. We remove the cross-verify query for clarity.}\label{fig:QA}
\end{figure}

Querying LLM with a single complex question can lead to incorrect responses~\cite{gpt_quesiton0}. Related study~\cite{gpt-chain} suggests that the Chain of Thought (CoT) reasoning method can enhance the LLM's comprehension of complex questions. In particular, we break down the rules for the documentation into a series of intermediate questions. As shown in Figure~\ref{fig:QA}, to check Rule 1 (i.e., \emph{Member Participation}), we raise three questions: \emph{Does the DAO support governance?}, \emph{Who can become a member of DAO?}, and \emph{Can members participate in governance?}. If all three questions are confirmed in the documentation, we determine that the rule is satisfied. For all 6 rules, we address them with a series of questions based on their detailed requirements from the DAO Model Law and merge similar queries to form a question chain, as depicted in Figure~\ref{fig:QA_chain}.


We utilize the ChatGPT model \emph{gpt-3.5-turbo-16k-0613} for classification. The prompt for each query is shown in the first box of Figure~\ref{fig:QA}.
We cross-verify the results with the following query: \emph{Your task is to check if the sentence content is mentioned in the document. Here is the sentence: [REASON]. Your answer format should be Result: Yes/No. The document is provided below: [DOCUMENT]}. If the documentation exceeds the token limitation, we partition it into segments, each consisting of 12,000 tokens. Every two segments share an overlap of 2,000 tokens.

\begin{table}[]
\caption{ Evaluation of checking whether the rule is mentioned in the documentation.}\label{tab:doc_manual}
\resizebox{0.99\columnwidth}{!}{%
\begin{tabular}{@{}lrrr||rrr@{}}
\hline
\multirow{2}{*}{Rule Name} & \multicolumn{3}{c||}{ChatGPT~\cite{chatgpt}} & \multicolumn{3}{c}{Claude~\cite{claude}} \\ \cmidrule(l){2-7} 
                           & Recall& Precision & F1-score  &Recall & Precision & F1-score      \\  \midrule
Member Participation                            & 0.69  & 0.95      & 0.80      &0.74     &   0.95    & 0.83     \\
Member Exit                                     & 0.00  & 0.00      & 0.00      &  0.00     & 0.00      &  0.00    \\
Voting Power                                    & 0.81  & 0.87      & 0.84      &  0.78     &    0.48   &  0.59    \\
Minority Protection                             & 1.00  & 1.00      & 1.00      & 1.00      &   0.50    &  0.66    \\ 
Governance Process Guide                        & 0.68  & 0.90      & 0.77      &  0.92     &   0.80    &  0.86    \\ 
Appointment of Guardian                         & 0.89  & 1.00      & 0.94      &   0.89    &    0.73   &   0.80   \\ \bottomrule
\end{tabular}
}
\end{table}



\noindent \textbf{Result.} The number of collected websites and documentation are shown in Table~\ref{tab:DAO_dataset}. We find that for all types of DAOs, only a small proportion of them, specifically 622 out of 16,427, provide their website. However, we discover that more than 460 of these websites are either offline or have expired domain names. Consequently, only 183 of these DAO websites are still operational.
We think that the cause might be the lack of maintaining a website, leading only to popular DAOs can create and sustain their websites. To validate our hypothesis, we examine the DAOs with a TVL exceeding 20 million dollars from CoinMarketCap~\cite{CoinMarketCap}. We find all of these 11 DAOs still maintain their websites. We subsequently analyzed if the low online website rate is due to DAOs being out-of-service. A DAO is considered out-of-service if it has not proposed any proposal within a year. We find that most DAOs (14,335 out of 16,427) are still active. However, only 183 of them have maintained their websites.





To evaluate the effectiveness of the ChatGPT, we randomly select 100 documentation and manually analyze whether each documentation satisfied each rule. The results are presented in Table~\ref{tab:doc_manual}. The results demonstrate that while the question chain can assist ChatGPT in accurately determining whether a certain rule is mentioned in the documentation, the length of the question chain can decrease the recall rate due to the false negatives from ChatGPT. In addition to ChatGPT, we also evaluated another LLM, Claude~\cite{claude}, for comparison. The evaluation, detailed in Table~\ref{tab:doc_manual}, reveals that Claude can produce results with a recall rate that matches or even exceeds that of ChatGPT for these questions. This suggests that both LLMs can provide sufficiently accurate outcomes. However, Claude produces more false positives than ChatGPT. The higher rate of false positives could be attributed to differences in training data or the possibility that Claude requires a different prompt structure compared to ChatGPT. Thus, we choose ChatGPT to measure the integrity of DAO documentation.

The results of each DAO's documentation and how they align with the rules set by the DAO Model Law are illustrated in Table~\ref{tab:doc_des}. Our findings show that none of the documentation from these DAOs adheres to all 6 rules. We discover that Rule2 \emph{Member Exit} is not mentioned in any of the DAO documentation. Further analysis of the DAO Model Law indicates that this rule acts more as a compliance standard than a practical guideline for DAOs. In practice, the removal of all DAO tokens belonging to a member is considered as the member's exit from the DAO by default. As for Rule3 \emph{Minority Protection}, only the DAO belongs to DAOhaus, has mentioned it in the documentation. After analyzing the DAOhaus~\cite{daohaus} platform, we find it incorporates the \emph{rage quit} procedure into their governance to ensure the protection of members with less voting power.

\begin{table}[]
\caption{ Integrity of documentation from ChatGPT.}\label{tab:doc_des}
\centering
\renewcommand{\arraystretch}{0.8}
\begin{tabular}{@{}lrrrrrr@{}}
\toprule
DAO Type & Rule1 & Rule2 & Rule3 & Rule4 & Rule5 & Rule6 \\ \midrule
XDAO & 1 & 0 & 0 & 0 & 2 & 0 \\
Aragon & 12 & 0 & 11 & 0 & 8 & 3 \\
Tally & 34 & 0 & 26 & 0 & 28 & 6 \\
DAOhaus & 1 & 0 & 0 & 1 & 0 & 0 \\
DAOstack & 1 & 0 & 1 & 0 & 1 & 0 \\
Self-developed & 3 & 0 & 3 & 0 & 3 & 2 \\ \midrule
Total & 52 & 0 & 41 & 1 & 42 & 11 \\
\bottomrule
\end{tabular}
\end{table}





\answer{2}{Of the 16,427 DAOs analyzed, only 622 provide their websites. Among these, only 183 DAOs continue to keep their websites operational, and none of these 183 DAOs provide sufficient documentation for the governance process.}

\section{Do existing proposals ensure consistency between descriptions and code?(RQ3)}\label{RQ3}

The governance process has become the target for attackers because the code within the proposal is controlled by members. As such, attackers can hide malicious code into the proposal with the intention of either gaining control over the DAO or transferring all the assets. In recent years, there have been numerous instances of governance attacks within DAOs, resulting in the loss of millions dollars~\cite{yam_attack, curve_attack, audius_attack, Fortress_protocol_attack, create2_attack, True_Seigniorage_Dollar_Attack, Pride_punks_attack, buildfinance_attck, yuan_attack, beanstalk_exploit}. To explore the security issues in the proposal, we first verify the immutability of the proposal code by ensuring the \emph{target address} in the proposal is open-source and is not created with the opcode $CREATE2$. Second, we check the consistency between the proposal description and code by verifying all code within the proposal is mentioned in the description.

\subsection{Immutability of Proposal Code}
To assess the immutability of the proposal code, we analyze the \emph{target address} within the proposal. The \emph{target address} refers to the contract to be called in the proposal code. It should be open-source so that members can examine the code logic within the target address. Apart from being open-source, the code logic in the \emph{target address} must also be immutable. As discussed in Section~\ref{lab:RQ1_3}, the EVM opcode $CREATE2$ allows arbitrary change to the code logic inside the \emph{target address} while maintaining the same address. If the proposal code lacks immutability, an attacker can arbitrarily modify the code even after the proposal has been approved. For instance, in the Tornado Cash Governance Attack~\cite{create2_attack}, the attacker initially proposes a benign proposal, which is subsequently approved. However, before executing the proposal, the attacker modifies the code within the proposal to transfer all assets.

\noindent \textbf{Approach.} To assess whether the target address in the proposal code is open-source, we follow the approach used in the previous study~\cite{NFT_measure}. We use APIs provided by blockchain scanners to check if the source code has been verified. We use the same method used in Section~\ref{lab:RQ1_3} to check the \emph{target address} is under the threat of the opcode $CREATE2$. We skip the \emph{target address} that belongs to the governance contract, as it has been evaluated in Section~\ref{lab:RQ1_3}.

\noindent \textbf{Result.} The results of the immutability of the proposal code are shown in Table~\ref{tab:target_address_result}. We discover that more than 99\% of the \emph{target address} in the proposal code are open-source. This suggests that the majority of proposals maintain the clarity of their proposal codes. Among the 1,172 closed-source contracts, we identify 24 addresses that have been used by members, as indicated by more than 500 transactions associated with these specific addresses. This implies that some members place their trust in these contracts despite the noticeable lack of transparency. Regarding $CREATE2$, although we do find some target addresses created in the $CREATE2$ chains, they can not destruct themselves and thus are not at risk of being mutated. However, the attacker can insert the $SELFDESTRUCT$ or $DELEGATECALL$ into the target address's code to make this potential threat feasible.

\begin{table}[!t]
\renewcommand{\arraystretch}{0.75}
\caption{ Result of the immutability of target address within proposal code.}\label{tab:target_address_result}
\resizebox{0.99\columnwidth}{!}{%
\begin{tabular}{@{}lrr||rr@{}}
\toprule
DAO Type  & Open-source  & Close-source  & By $CREATE2$  & Can $SELFDESTRUCT$  \\ \midrule
XDAO      & 96,588  & 568         &  14            & 0\\
Aragon    & 14,843 & 3             & 0 & 0\\
Tally     & 4,468 & 540          & 124&0 \\
DAOhaus   & 1,704 & 56            & 2 &0\\
DAOstack  &  290 & 0            & 0 &0 \\
Self-developed & 1,161 & 5          & 0 &  0  \\ \midrule
Total  &119,054 & 1,172& 140 & 0 \\
\bottomrule
\end{tabular}
}
\end{table}

\subsection{Consistency between Description and Code}
The proposal description should align with the proposal code, ensuring members are fully informed about the proposal's intent and allowing them to cast an informed vote. If there's a discrepancy in consistency, it could jeopardize the correctness of the final voting result.
To verify the consistency between the proposal description and code, we employ a framework containing three components: description intention extractor, code action extractor, and inconsistency detector. Within the description intention extractor, we extract the \emph{description intention}—identified as \emph{(action, target object, parameter)}—from the proposal description, which outlines the functions to be called or not called in the proposal code. The code action extractor is used to gather the \emph{code action} (as detailed in Table~\ref{tab:proposal_code_fact}) from the proposal code, which shows the actual functions to be executed. Finally, we assess the consistency between the \emph{description intention} and the \emph{code action} in the inconsistency detector.

\subsubsection{Description Intention Extractor} 
\begin{figure}
\begin{center}
\includegraphics[width=0.99 \columnwidth]{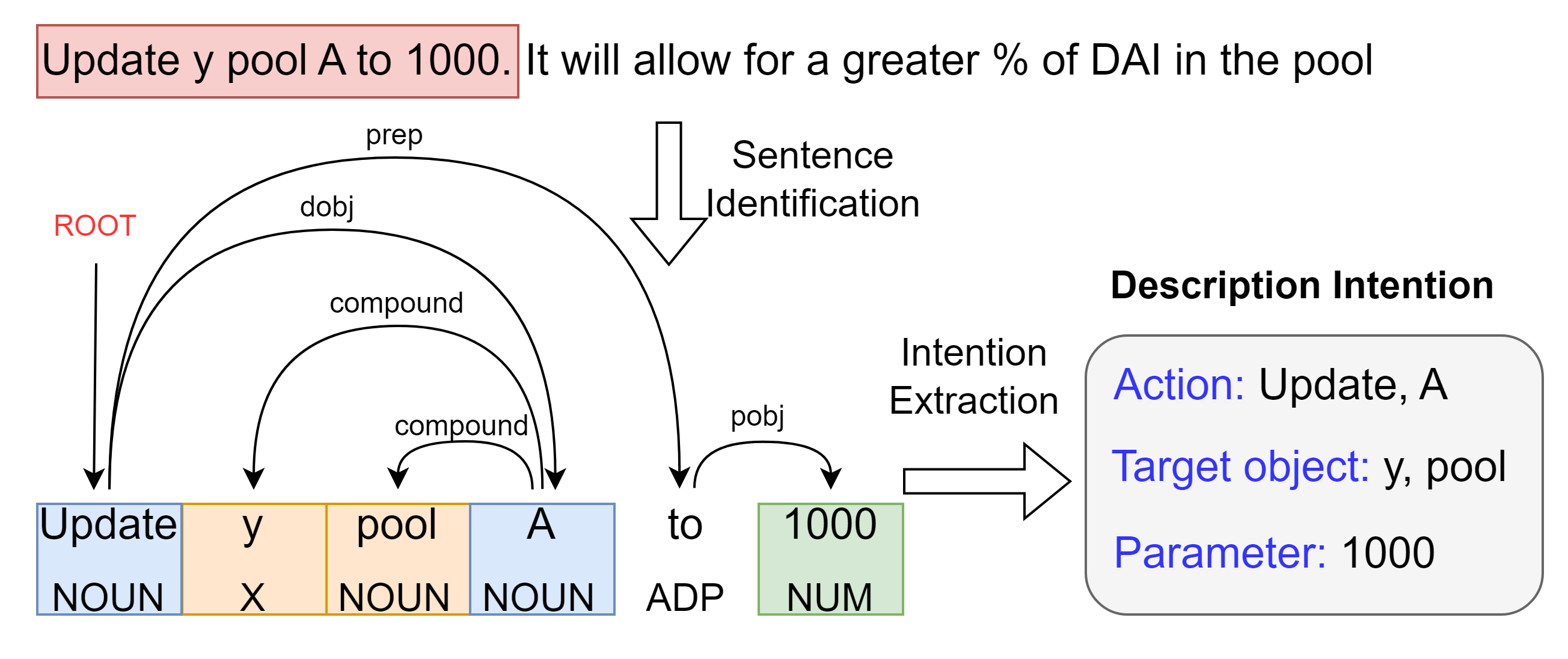}
\end{center}
\caption{ Example of extracting the \emph{description intention} from the proposal description.}\label{fig:DI}
\end{figure}

The \emph{description intention} is represented as a tuple \emph{(action, target object, parameter)}. The \emph{action} refers to the function name to be performed by the proposal code (e.g., transfer, update, approve), \emph{target object} is the target of the function call, and \emph{parameter} denotes the detailed parameters used by the \emph{action}. We adopt a two-step process to extract the \emph{description intention} from the proposal description. First,  We identify all the code-related sentences that describe the function calls in the proposal code. After that,  we extract the \emph{description intention} from these code-related sentences based on their grammatical structures. The example procedure of the description intention extractor is shown in Figure~\ref{fig:DI}. The sentence in the red box is identified as code-related. Subsequently, during the intention extraction, the code-related sentence undergoes parsing to form the corresponding semantic dependency parse tree. The \emph{description intention} is then extracted based on the part-of-speech tags and syntactic dependencies in the parse tree.

\noindent\textbf{Sentence identification.} We apply the NLTK~\cite{nltk} to split the proposal description into individual sentences. In order to identify code-related sentences that describe the code, we use a fine-tuned BERT~\cite{bert} for the sentence classification task. Due to the absence of a dataset for code-related sentences in DAO proposals,  we created a dataset comprising 2,200 sentences randomly extracted from proposals. We select 2,000 sentences from this dataset to fine-tune the BERT model. The remaining 200 sentences are used to evaluate the performance of the find-tuned BERT. We manually annotate each sentence to indicate whether it describes the proposal code. The evaluation of sentence identification is shown in Section~\ref{RQ3_E1}.


\noindent\textbf{Intention extraction.} To extract the \emph{description intention} from code-related sentences, we first use Spacy~\cite{Spacy} to generate a syntactic dependency parse tree and assign part-of-speech (PoS) tags to each token within the sentence. The $action$ is identified by the token that is labeled as $Root$ in the PoS tag. Its lemma either exists in our verb list\footnote{\parbox{\columnwidth}{https://drive.google.com/file/d/1I1mPkZMohjC8vINL9JvJSoN8SoymDTRO}}, or it aligns with synonyms of words within our verb list, as determined by the synonyms database~\cite{synonyms_db}. 
Additionally, the token that has a \emph{direct object (dobj)} relationship with the $Root$ token is also identified in the $action$. The \emph{target object} is identified by tokens that have a $compound$ relationship with the $action$ tokens. Lastly, the $parameter$ is identified by the rest tokens with PoS tags such as $NOUN$, $NUM$, $PROPN$, or $X$. As shown in Figure~\ref{fig:DI}, the $action$ is highlighted in the blue box, the \emph{target object} in the yellow box, and the $parameter$ in the green box. We also identify whether the \emph{description intention} originates from negative or positive sentences. To identify these negative sentences, we utilize the BERT to determine whether the code-related sentence is positive or negative. When extracting from these negative sentences, we assign a \emph{Negative} tag to the \emph{description intention}.

\subsubsection{Code Action Extractor}
\begin{figure}
\begin{center}
\includegraphics[width=0.85 \columnwidth]{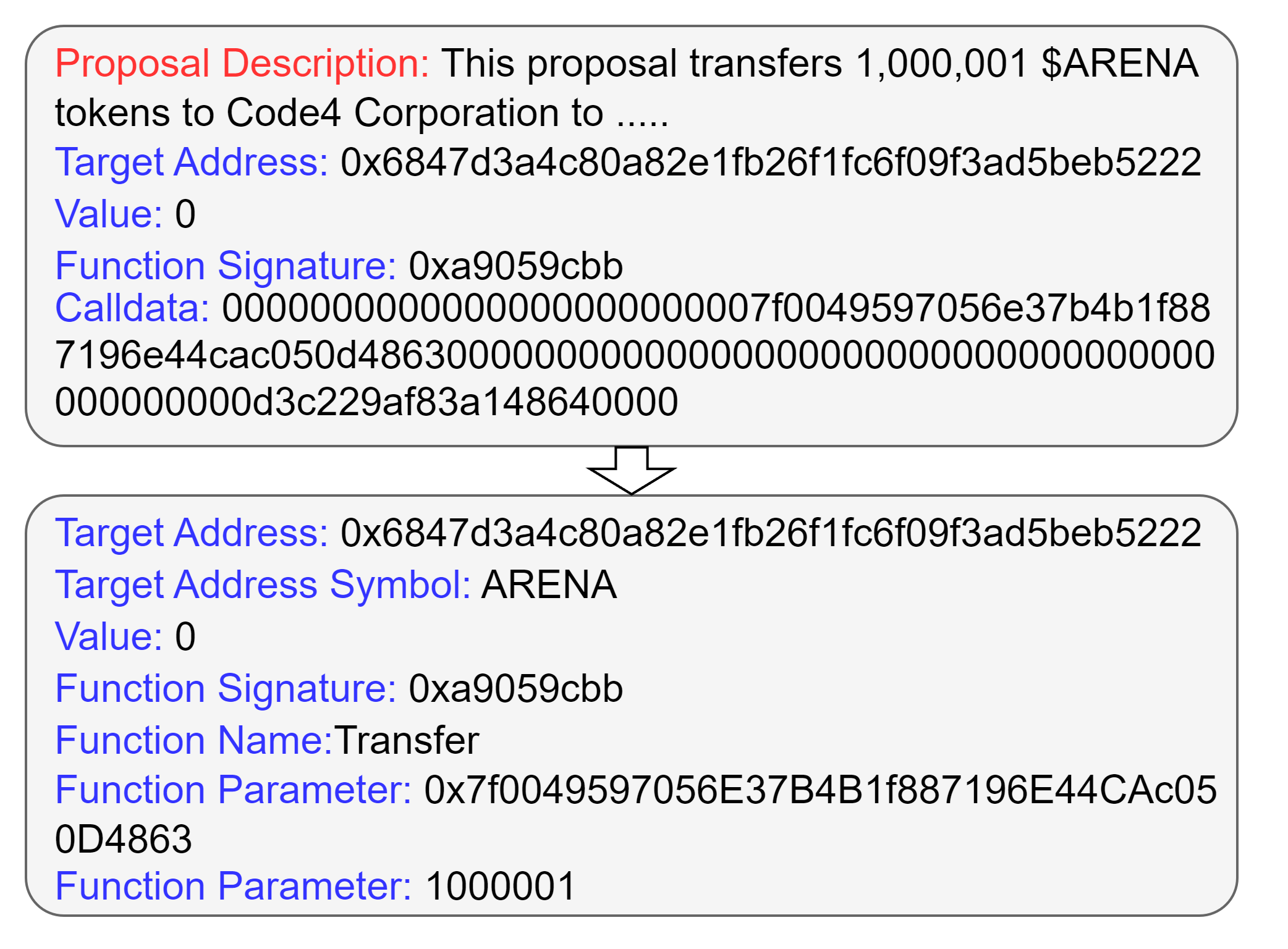}
\end{center}
\caption{The example illustrating the extraction and enhancement of proposal code into \emph{code action}.}\label{fig:CA}
\end{figure}

The code action extractor's purpose is to extract the proposal code and enrich its content, resulting in the \emph{code action} as illustrated in Table~\ref{tab:proposal_code_fact}.
Given that the proposal code is in bytecode format, verifying its consistency with the \emph{description intention} could result in false negatives. For instance, in Figure~\ref{fig:CA}, the proposal description outlines its object as \emph{transfer of ARENA tokens}. It is challenging to determine if the code matches the description directly from the bytecode. To address this, we transform the proposal code into \emph{code action} to add natural language information. 

Since the proposal code only contains the $target\ address$, $value$, $function\ signature$, and $calldata$ from the proposal code, the rest part of \emph{code action} needs to be enhanced based on these data. The \emph{target address symbol} can be determined by checking the contract address in the public name tag or the function named \emph{symbol()} in the contract.
To determine the $function\ name$, we attempt to find it either from the contract ABI~\cite{contract_ABI} of the target address or from the Ethereum Signature Database~\cite{4bytes}. We obtain the contract ABI from the source code of the \emph{target address} via blockchain scanner.
If the target address is closed-source, we turn to the Ethereum Signature Database~\cite{4bytes}—the largest database mapping \emph{function signature} back to \emph{function name}—for querying the $function\ name$.
Once the $function\ name$ is extracted, we can decode the $calldata$ into \emph{function parameter}, due to the $function\ name$ containing the sequence and types of each parameter. If we are unable to locate the corresponding information, we will leave it empty.

\begin{table}[]
\caption{ Component of \emph{code action.}}\label{tab:proposal_code_fact}
\renewcommand{\arraystretch}{0.75}
\resizebox{0.99\columnwidth}{!}{%
\begin{tabular}{@{}ll@{}}
\toprule
 Name & Explanation \\ \midrule
    Target Address & Contract address to be called      \\ 
    Target Address symbol & Contract address name in natural language  \\
    Value & Value to transfer  \\
    Function Signature & 4 bytes of function ID   \\ 
    Function Name & Function name in natural language  \\
    Function Parameter & Parameter of the function  \\
    \bottomrule
\end{tabular}%
}
\end{table}

\subsubsection{Inconsistency Detector}
We determine the following 5 types of inconsistency between the \emph{description intention} and \emph{code action}.

\noindent\textbf{Lack of \emph{description intention}.} We extract the \emph{code action} from the proposal code, but can not find any \emph{description intention}.

\noindent\textbf{Lack of \emph{code action}.} We extract the \emph{description intention} from the proposal description but fail to find any \emph{code action}.

\noindent\textbf{Incomplete function.} This type arises when a function present in the \emph{code action} is not referenced in the \emph{description intention}. Our analysis involves verifying whether every \emph{Target Address Symbol} and \emph{Function Name} from the \emph{code action} are explicitly mentioned in the corresponding \emph{Target object} and \emph{Action} within the \emph{description intention}.
We compare the semantic similarity between the combination of \emph{function name} and \emph{target address symbol} from the \emph{code action} with the $action$ in the \emph{description intention}. If the semantic similarity score exceeds the threshold, we believe the function is mentioned in the description since the description describes the function in a similar semantic meaning. We utilize the Sentence-BERT model~\cite{sentence-bert} to convert the corresponding sentences into vectors and calculate the cosine similarity between these vectors. The threshold value is set at $0.75$, following the official Sentence-BERT examples~\cite{Sentence-BERT_Example}.




\noindent\textbf{Incomplete parameter.} We verify that every \emph{Function Parameter} listed in the \emph{code action} is referenced in the \emph{Parameter} section of the \emph{description intention}. We assess parameters of types: address, number, and byte, as well as their corresponding lists.
For the address type, we first retrieve its name using the method described in information enhancement. If we can extract the address name, we verify whether the name appears in the $parameter$ from the \emph{description intention}. If we cannot find the name, we directly check whether the address, in hexadecimal format, is mentioned.
For the number type, we verify whether the number is shown in the \emph{description intention}. If the target contract is an ERC-20, we divide the value by its decimals~\cite{erc_20}.
For the byte type, since it can be a hexadecimal representation of text, we check both the original content and the decoded text.

\noindent\textbf{Incorrect proposal.} The function in the \emph{code action} is mentioned by a \emph{description intention} tagged as \emph{Negative}. We apply the same method used in the incomplete function to verify the function is mentioned in the \emph{description intention}.


\subsubsection{Evaluation of Sentence Identification}\label{RQ3_E1}
The evaluation of fine-turned BERT demonstrate that it achieves the precision of 0.97, recall of 0.85. False positives occur when sentences aim to explain the functionality of the proposal code, rather than describing the actual proposal code itself. False negatives are generated when sentences use only abstract nouns to describe the proposal code.




\subsubsection{Evaluation of Incomplete Function and Parameter}
To evaluate the performance of incomplete function and parameter, we randomly select 1,500 functions, which contain 3,122 parameters. We manually label each of the functions or parameters as incomplete. The incomplete function achieved a precision of 0.81 and a recall of 0.87. The false positive is generated because the BERT can not correctly handle the relationship between some words with similar semantic meanings, especially for the description having only a few noun words. The false negative is due to the different meanings of the words between the real world and blockchain network. For the incomplete parameter, we achieve a precision of 0.83 and a recall of 0.88. The false positives are caused when the proposal description uses URLs to outline its function parameters. The false negatives are caused by the misleading name of the parameter.

\subsubsection{Result of Inconsistency between Proposal Description and Code}
The results of the 5 types of inconsistency are shown in Table~\ref{tab:incomplete_result}. We exclude proposals from XDAO, as this platform does not support proposals that include descriptions. Instead, it only requires members to submit code directly to the governance contract as a proposal, which could interfere with the analysis results.
Our results suggest that members currently do not pay sufficient attention to proposals. Of the 25,051 proposals analyzed, 19,148 either lack a description of the proposal code or only contain a description without corresponding code. Furthermore, among the 5,903 proposals that do include both a description and code, 3,830 are found to be incomplete, either lacking an explanation about the functions or detailed parameters in the functions.
This explains why attackers frequently target proposals in the governance process, as approximately 90 percent of proposals fail to provide a clear and consistent description of the code for the members.


\begin{table}[]
\caption{ Result of consistency between description and code. The \emph{description intention} is short for $DI$ and \emph{code action} is short for $CA$.}\label{tab:incomplete_result}
\renewcommand{\arraystretch}{0.85}
\resizebox{1.0\columnwidth}{!}{%
\begin{tabular}{@{}lrrrrr||r@{}}
\toprule
Consistency Type         & Aragon    & Tally & DAOhaus   & DAOstack  & Self-developed& Total\\ \midrule
Normal                   & 776       &  977  &  24        & 684       & 75            & 2,536\\
Lack of $DI$             & 14,601    & 743   & 674      & 30        & 449           & 16,497\\
Lack of $CA$             & 680       & 632   & 0         & 1,296     & 43            & 2,651\\
Incomplete               & 119       & 2,087 &  1,014      &  170      &440           &3,830\\
-- Function              & 37        & 1,848 &  894      & 110       &276            & 3,165\\
-- Parameter             & 215       & 9,186 & 1,644       & 356       & 544          & 11,945\\ 
Incorrect                & 0         &  0    &  0        & 0         & 0             & 0\\\midrule
Proposal                 & 16,149    & 3,953 & 1,712      & 2,190     & 1,047           & 25,051\\
Function                 & 14,693    & 6,498 & 2,457       & 210       & 1,361          & 25,219\\
Parameter                & 32,162    & 12,489& 1,981       & 533       & 816           & 47,981\\ \bottomrule
\end{tabular}
}
\end{table}

\subsubsection{Real-World Attack Cases}
To assess whether our approach is capable of detecting real-world malicious proposals, we have gathered reports of DAO governance attack cases from the following sources: Slowmist~\cite{slowmist}, CryptoSec~\cite{cryptocec}, Rekt~\cite{rekt}, and Twitter~\cite{twitter}. We total collected 11 DAO governance attack cases~\cite{yuan_attack, True_Seigniorage_Dollar_Attack, buildfinance_attck, beanstalk_exploit, audius_attack, Venusattack, yam_attack, Fortress_protocol_attack, Swerveattack, create2_attack, Atlantis_attack, Indexed_Finance_attack, BIGCAP_attack}. 
Upon examining these malicious proposals with our approach, we identified all 13 proposals as 8 malicious proposals due to lack of \emph{description intention}, 3 proposals due to incomplete function, 2 proposals with incomplete parameter, and 1 proposal is subjected to mutability of proposal code.

\begin{table}[]
\caption{Classification of the collected real-world governance attack incidents. The \emph{description intention} is short for $DI$ and \emph{code action} is short for $CA$.}\label{tab:read_attacks}
\renewcommand{\arraystretch}{0.85}
\resizebox{1.0\columnwidth}{!}{%
\begin{tabular}{@{}lrlrl@{}}
\toprule
Incidents & Date & Result & Expect Lost & Proposal Consistency  \\ \midrule
True Seigniorage Dollar~\cite{True_Seigniorage_Dollar_Attack} & Mar 2021 & Successed & \$16K & Lack of $DI$  \\
Yuan~\cite{yuan_attack} & Sep 2021 & Successed & \$250K & Lack of $DI$  \\
Venus~\cite{Venusattack} & Sep 2021 & Successed & \$250K & Lack of $DI$ \\
Build Finance~\cite{buildfinance_attck} & Feb 2022 & Successed & \$470K & Lack of $DI$  \\
Fortress Protocol~\cite{Fortress_protocol_attack} & May 2022 & Successed & \$3M & Incomplete Parameter \\
Beanstalk~\cite{beanstalk_exploit} &Apr 2022 & Successed & \$182M & Incomplete Function  \\
Audius~\cite{audius_attack} & Jul 2022 & Successed & \$1.1M & Lack of $DI$  \\

YAM~\cite{yam_attack} & Jul 2022 & Blocked & \$2.1M & Incomplete Function  \\

Swerve Finance~\cite{Swerveattack} & Mar 2023 & Successed & \$1.3M & Lack of $DI$ \\
Tornado Cash~\cite{create2_attack} & May 2023 & Successed & \$2M & Code Mutability  \\
Atlantis Loans~\cite{Atlantis_attack} & Jun 2023 & Successed & \$1M & Lack of $DI$  \\
BIGCAP~\cite{BIGCAP_attack} & Sep 2023 & Blocked & \$45K & Incomplete Function \\
Indexed Finance~\cite{Indexed_Finance_attack} & Nov 2023 & Blocked  & \$158K &  Lack of $DI$\\

\bottomrule
\end{tabular}
}
\end{table}





\answer{3}{Of the existing 25,051 proposals, 22,515 of them (approximately 90\%) fail to provide a consistent description and code for their members. Moreover, 16,497 of these proposals do not provide any description of their intended purpose.}

\section{Threat to Validity}
\noindent \textbf{Limitations of complete DAO data.} To counteract the threat of complete DAO data, we utilized several methods. First, we collect DAO data sources not only based on previous studies~\cite{dao-analyzer, DAO-voting-platform}, but also from famous industry datasets, such as DeepDAO~\cite{deepDAO}. Second, we collect DAOs not only from Ethereum but also from 8 different blockchains. Third, we collect DAO data from both websites and blockchains to further make the dataset more complete. Hence, our dataset, encompassing over 37,000 DAOs, 600 websites, and 100,000 proposals, represents the most comprehensive DAO dataset to date. The results derived from this dataset can be considered a representation of the entire DAO ecosystem.

\noindent \textbf{DAOs from non-EVM-compatible chains.} According to statistics from DefiLlama~\cite{chain_top}, EVM-compatible chains currently dominate the blockchain. They account for over 85\% of the TVL across all blockchains. Therefore, we primarily apply our approach to EVM-compatible chains. However, apart from the immutability of contracts, our methodology and obtained insights do not exclusively rely on features specific to EVM. Therefore, our approach can be applied to non-EVM-compatible chains as well.

\noindent \textbf{Off-chain governance DAOs.} In off-chain governance, the governance process takes place on the website, where members submit proposals and cast their votes. The execution of these proposals is carried out by the DAO developers rather than being automatically triggered by smart contracts~\cite{two_governance_type,two_governance_type-2}. According to the definition of DAOs provided by Ethereum~\cite{ETH_WP} and the DAO Model Law~\cite{model_law}, DAOs must be governed by smart contracts. Therefore, off-chain governance DAOs fall outside of our scope.

\section{Implications and suggestions.}
Based on our research findings, we recommend that DAO platforms should ensure that all DAOs established on their platforms adhere to the principles of decentralized governance rather than permitting developers to optionally support it. With regards to developers, they should be obliged to disclose all privileged addresses to their members or, alternatively, mandate that all privileged functions be owned by the governance contract. Furthermore, they should provide complete documentation to aid members in participating in the governance process. Blockchain scanners, such as Etherscan, should label contracts that are deployed by the opcode \emph{CREATE2}. In response to the observed inconsistencies in proposals, we suggest the DAO should force the consistency between proposal description and tools and develop tools that can automatically supplement proposal descriptions with missing proposal codes and explanations.
\section{Related Work}


\noindent\textbf{DAO.}
Recent research on DAO focuses on the DAO activity analysis~\cite{dao_gas_price, DAO_overview, Platform_based_DAO, DAO_number_changes, influence_the_token_price, vulnerableDAO}, DAO definition and application~\cite{ choose_dao, dao_survey}, and DAO governance method~\cite{novel_on_chain, 2018chain_governance, multi_chain_token}. However, they do not concentrate on the security aspects of DAO governance. As for empirical studies that do focus on security within DAO governance: Feichtinger~\emph{et al.}\cite{empirical_on-chain} provided analysis on 21 on-chain governance DAOs, specifically focusing on the voting process within the governance procedure. Fritsch~\emph{et al.}\cite{empirical_voting} focused on the distribution of voting power among three popular DAOs: Compound, Uniswap, and ENS. Sharma~\emph{et al.}\cite{empirical_practice} analyzed the existing centralized risk of 10 existing DAOs and the corresponding members voting behaviors. Wang~\emph{et al.}\cite{DAO-voting-platform} analyzed the design principles of DAOs from off-chain voting platform Snapshot. Liu~\emph{et al.}\cite{rw_voting} focused on voting behavior in DAO governance. Dotan~\emph{et al.}\cite{vulnerableDAO} disclosed the centralized voting nature of four DAOs and explained the existing governance attack incidents. The above research primarily focused on partial aspects such as voting within the DAO governance framework, and their datasets are limited, no larger than 1,000 DAOs. Our methodology analyzes the security issues across both on-chain and off-chain parts of the governance framework. The security threats we studied have not been explored in previous research.

\noindent\textbf{Smart contracts analysis.}
Smart contracts have gained popularity for facilitating trustless code execution on the blockchain. However, with the increasing usage of smart contracts, they have become targets for attacks. Numerous tools have been developed for the analysis of smart contracts. Some notable examples include Mythril~\cite{Mythril}, Manticore~\cite{Manticore}, and Oyente~\cite{Oyente}. Pied-Piper~\cite{caller2} proposed a hybrid analysis method that combines datalog analysis and directed fuzzing to detect potential backdoor threats in ERC token contracts in order to enhance smart contract security. Beyond the direct analysis of bytecode, binary lifter tools such as Gigahorse~\cite{gigahorse} transform the bytecode into a higher-level, function-based, three-address representation. Our method targets the detection of security issues within governance contracts and can be integrated with existing tools to enhance the security of dApps.


\noindent\textbf{Consistency between code and natural language description.} 
The consistency between the code and natural language description has been well-studied~\cite{ppchecker, pos_tag, rw3, rw3_0, rw3_1, rw3_2}. They primarily concentrate on Java code and API documentation, which are well-written and focused on describing code behavior. DocCon~\cite{doccon} detects inconsistencies between documentation and the corresponding code for Solidity smart contract libraries. Compared with Doccon, our method targets different research questions. Our natural language description comes from proposal description, which lacks structured information such as tags in the comments or API document. Additionally, the proposal description encompasses a broader scope instead of only describing the code behavior. The code in our method is the bytecode, not the Solidity source code, which lacks code information like variable name. Furthermore, our code size is extremely limited, containing only several bytes and the function call parameters rather than the full code logic.

\section{Conclusion}
In this paper, we conduct a comprehensive study of the issues in the DAO governance process components. We construct the dataset contains 16,427 DAOs, 183 documentation, and 122,307 proposals across 9 different blockchains. Then we apply our novel methods to automatically identifying issues within these components.
Our analysis of the governance contract shows there are 665 DAOs the privileged functions within the governance contract can be controlled by unknown entities that might be used by developers to intervene the governance process. In terms of documentation, 99\% of DAOs fail to provide documentation related to the governance process. Analysis of proposals reveals that 90\% proposals fail to provide a detailed description explaining the proposal code. 
\bibliographystyle{ieeetr}
\bibliography{./sample-acmsmall-conf.bib}


\end{document}